\documentclass[preprint, 3p]{elsarticle}

\usepackage{graphics}
\usepackage{subcaption}
\usepackage{rotating}
\usepackage{bm}
\usepackage{mathrsfs}
\usepackage{amssymb,amsmath}
\usepackage{float}
\usepackage{siunitx}
\usepackage{blindtext}
\usepackage{mathtools}
\usepackage{xcolor}
\usepackage{csquotes}
\usepackage{tikz}
\usepackage{float}
\usepackage{textcomp}
\usepackage{soul} 
\usepackage{adjustbox} 

\usepackage[colorlinks,linkcolor=red,citecolor=blue,urlcolor=blue]{hyperref}

\biboptions{sort&compress} 

\usepackage{amssymb}
\usepackage{pifont}
\usepackage{lineno}
\usepackage[normalem]{ulem}

\newcommand{\fone}{f^{(1)}}

\newcommand{\uu}{\bm{u}}

\newcommand{\fzero}{f^{(0)}}

\newcommand{\ux}{\bm{x}}
\newcommand{\uxi}{\bm{\xi}}

\newcommand{\uP}{\bm{P}}
\renewcommand{\H}{\mathcal{H}}
\newcommand{\al}{\alpha}
\newcommand{\ua}{\bm{a}}

\title{Palabos Turret: A Particle-Resolved Numerical Framework for Settling Dynamics of Arbitrary-Shaped Particles}

\author[aff1,aff2]{Taraprasad Bhowmick\corref{cor1}\fnref{eqcont}}
\author[aff3]{Jonas Latt\corref{cor1}\fnref{eqcont}}
\author[aff1]{Yong Wang\fnref{eqcont}}
\author[aff1]{Gholamhossein Bagheri}

\address[aff1]{Max Planck Institute for Dynamics and Self-Organization, 37077 Göttingen, Germany}
\address[aff2]{University of G\"ottingen, 37077 Göttingen, Germany}
\address[aff3]{University of Geneva, 1211 Genève 4, Switzerland}

\cortext[cor1]{Corresponding authors: taraprasad.bhowmick@ds.mpg.de, jonas.latt@unige.ch}
\fntext[eqcont]{Contributed equally to this work}

\date{\today}

\begin{document} 

\begin{abstract}
Particles transported in fluids are everywhere, occurring for example in indoor air, the atmosphere, the oceans, and engineering applications. In this study, a novel three-dimensional numerical framework -- the Palabos Turret is presented, which allows fully resolved simulations of the settling dynamics of heavy particles with arbitrary shapes over a wide range of particle Reynolds numbers. The numerical solver is based on the lattice Boltzmann method utilizing immersed-boundary approach and a recursive-regularized collision model to fully resolve the particle-fluid interactions. A predictor-corrector scheme is applied for the robust time integration of the six-degrees-of-freedom (6DOF) rigid-body motion. Finally, the multi-scale nature arising from the long free-fall distances of a particle is addressed through a dynamic memory allocation scheme allowing for a virtually infinite falling distance. This solver allows for the simulation of particles of any arbitrary shape. The proposed framework is validated using the analytical and experimental data of freely-falling spheres, ellipsoids, and an irregular particle in a wide range of Reynolds numbers between $5\times10^{-1}$ and $4\times10^4$. For different Reynolds numbers and particle shapes considered, the Palabos Turret shows excellent agreement compared to theoretical and experimental values with a median relative deviation of $\pm1.5\%$ and a maximum deviation of $\pm5\%$. The Palabos Turret enables an in-depth analysis of the translational and rotational dynamics of particles with complex geometries.
\end{abstract}


\maketitle

\section{Introduction}\label{sec:intro}
Solid particles moving in fluids are ubiquitous, from indoor air to the atmosphere and ocean to engineering applications. Such particles are almost universally non-spherical. Examples of such particles are volcanic ash \cite{Rossi_2021}, dust \cite{van_der_Does_2018}, pollen \cite{vanHout:2004}, ice crystals \cite{Heymsfield_1973}, marine snow \cite{Trudnowska_2021} and microplastics \cite{Xiao_2023}, which mostly range in size from a few tens of micrometres to several centimetres (for the sake of brevity we exclude very small and large particles here). Their dynamics are influenced by their material properties such as density, size and shape, as well as by the properties of the ambient fluid and the flow characteristics. There is a long list of theoretical, experimental and numerical studies in the literature dealing with this topic \cite{willmarth1964steady,Khayat_Cox_1989,Newsom_Bruce_1994,Bagheri_Bonadonna_2016,Gustavsson_2019,ouchene2020numerical,Alipour_2021,Tinklenberg_2023,Tatsii_2023,bhowmick_PRL_2024}. While experimental studies can provide the ground truth to guide and validate models, a reliable simulation can provide much more information about the particle dynamics that is hardly achievable in experiments. In recent years, several numerical methods have been proposed, which have demonstrated their capability in capturing some of the essential physical phenomena of such processes with high efficiency and accuracy \cite{Hou_2012,Mittal_2005,Tenneti_2014,Griffith_2020,MA_2022}.

The particle dynamics is usually characterised by the Reynolds number of the particles $\mathrm{Re} = v_\mathrm{T} d / \nu$, which is defined as the relative velocity of the particles (here terminal velocity $v_\mathrm{T}$) multiplied by a characteristic length of the particles (here particle diameter $d$) and divided by the kinematic viscosity of the fluid $\nu$. In a turbulent flow, other dimensionless parameters also play a role. 
However, even in a laminar flow, it is not only the Reynolds number of the particles that is important, but also the ratio of the particle density to the fluid density. It has already been shown that for a $\mathrm{Re} >$ 100, the terminal velocity increases with the density ratio \cite{Bagheri_Bonadonna_2016,Tinklenberg_2023}. Recently, it was also found that the angular dynamics of particles with Reynolds number between 1-10 is also significantly influenced by the density ratio \cite{bhowmick_PRL_2024}. 
If the particle is considered rigid, its time evolution can be described by a model with six-degrees-of-freedom, or `6DOF' for short, corresponding to three degrees of freedom for the position and three degrees of freedom for the orientation \cite{Thornton_2004}. The accuracy of the equations of motion relies on the computation of the total force and moment-of-force, which is obtained by integrating over an arbitrary particle shape \cite{Greenwood_1988}.

A numerical framework is generally tailored for specific Reynolds numbers and particle to fluid density ratio regimes. Abundant literature exists for numerical simulation of particle suspensions at low Reynolds numbers and with density ratios close to 1 \cite{Parsa_2012,Auguste_2013,Voth_Soldati_2017,Sheikh_2020}.
These simulations have wide-ranging applications including sedimentation processes for particles immersed in fluids, where the particle density places them in a close-to-buoyant regime.
To name another example, numerical simulation has made significant progress in representing blood components, including red blood cells or platelets, at the cellular level for biomedical process descriptions \cite{Almomani_2008,Freund_2014,YE_2016}.
In such cases, the use of a solver capable of capturing fluid-structure interaction is necessary due to the deformable nature of the blood cells.
The methods commonly rely on an immersed-particle philosophy, in which the fluid mesh is stationary, and the mesh points are allocated both outside and inside the particle. The common implementations of this philosophy are the saturated-cell approach, which relies on a computation of the solid fraction across the interface of the moving particle \cite{RETTINGER_2018,RETTINGER_2022}; or the immersed-boundary approach \cite{Mittal_2005,Griffith_2020}, which simulates a fluid equation inside the solid particle as well. Both approaches avoid the need for costly remeshing of fluid domain surrounding the moving particles.

One of the main difficulties in the experimental or numerical study of freely-falling particles is their multi-scale nature. For this problem, the length scales range from the size of the particle (e.g. a few hundred micrometres to centimetres) to the vertical drop distance which is between ten to thousand times the particle size to reach the terminal velocity, when the particle settles with a preferred orientation. When no preferred orientation is present, a large vertical drop distance is also necessary to collect sufficient statistics on the pseudo-terminal settling velocity of the particle. Together with the need for high resolution to correctly resolve the shape features of the particle and the particle-fluid interactions, this problem can be fundamentally intractable given current computational power. A common workaround involves restricting the effective numerical domain size, but in this case, the proximity to a no-slip wall representing the fixed reference frame (the ground) may lead to subsequent numerical and physical errors.

In this work, we introduce a novel numerical framework to model particle free-fall that addresses these issues. Our framework is developed on the immersed-boundary approach in the lattice Boltzmann modeling of rigid solid motion. This framework, referred hereafter as the `Palabos Turret', is an extension of the Palabos library~\cite{latt2021palabos} -- a general-purpose computational fluid dynamics (CFD) code with a kernel based on the lattice Boltzmann method (LBM). The numerical stability of the flow is improved using the recursive-regularized collision model \cite{coreixas2018high,coreixas2019comprehensive}. This model also includes a subgrid-scale filtering capability, achieved by projecting kinetic variables onto a Hermite-tensor subspace. This collision model therefore not only maintains stability but also ensures sufficient accuracy during local direct numerical simulation (DNS) to large-eddy simulation (LES) transition, which is expected to take place frequently in this flow regime. It will be demonstrated that the Palabos Turret allows for the retrieval of fluid flow values with high precision to determine the force and moment of force acting on the surface of the immersed particle.

Owing to the adopted velocity discretization scheme, the LBM is traditionally limited to the use of structured meshes with uniform, cubic cells. While this limitation is typically addressed by applying mesh refinement on a hierarchical, octree-like mesh structure, such an approach is not employed in the Palabos Turret, which instead uses a fully uniform mesh. One of the consequences of the uniform mesh structure is that the mesh is not body-fitted for any object that has a non-rectangular shape or is moving. As a workaround, the LBM community has a well-established a set of tools for accurately resolving boundary conditions for the moving, immersed objects.  Earlier methods, which were developed by Hu et al. (1992) \citep{hu:92}, Feng et al. (1994) \citep{feng:94.1,feng:94.2}, Ladd (1994) \citep{ladd:94.1,ladd:94.2}, Aidun \& Lu (1995) \citep{aidun:95}, and Johnson \& Tezduyar (1995) \citep{johnson:95}, applied a staircase pattern to represent a particle boundary that followed the shape of the fluid solver's mesh. They required a large particle size compared to the computational grid or mesh spacing to achieve stable and consistent results, or they needed to frequently remesh the fluid domain around the moving particles. Recently, a significant amount of research has been conducted to expand both conventional CFD approaches and the LBM in order to achieve a more precise representation of the surfaces of moving particles in fluids, without increasing the resolution of the computational domain \citep{glowinski:99, verberg:00, verberg:01, takagi:03, zhang:03, feng:04, li:04, strack:07, hoelzer:09, guo:11, wang:13}. In separate research endeavors, the LBM community has developed the ability to accurately represent curved boundaries over time \citep{ginzburg2023unified}.
In this work, we concentrate on a multi-directional forcing approach, adapted for the LBM. Such approach has previously been used to model particles with density ratios close to unity \cite{thorimbert2018lattice}. Our present work extends this approach to high density ratios and to particles of arbitrary shape. For this, the boundary representation is complemented with a time-integration scheme for the 6DOF of a rigid particle, leading to a complete computational framework for immersed particle motion.

The rest of this paper is divided into three parts.
Section \ref{sec:numerical method} outlines the details of the Palabos Turret for the particle free-fall experiment, which includes the recursive-regularized LBM for the fluid equations, the immersed boundary condition for the fluid-particle coupling, and a time-staggered predictor-corrector method for the time integration of the particle motion. To allow for a long free-fall distance without spending an excessive amount of memory, a dynamic memory allocation strategy is utilized. 
In Section \ref{sec:results}, after domain and mesh convergence tests, the validity of the Palabos Turret is firstly confirmed in two simple cases, the time-dependent behavior of a sphere experiencing rotation-free free-fall and a displacement-free rotational motion. The proposed Palabos Turret is then used to investigate different scenarios with freely-falling particles, such as spheroidal and irregular volcanic (lapilli-size) particles in a wide range of Reynolds number $\sim$ 0.5 - 40000 and a particle to fluid density ratio of $\sim$ 1000. The numerical results show a high degree of agreement with the analytical solutions and the experimental observations, validating the proposed Palabos Turret. Finally, a conclusion is drawn in Section \ref{sec:conclusion}.

\section{The Palabos Turret}\label{sec:numerical method}
\subsection{Lattice Boltzmann method}
In the present work, the air is modeled as an incompressible fluid, as governed by the incompressible Navier-Stokes equations:

\vspace{-5mm}
\begin{align}
\partial_t{\bm u} + \left({\bm u}\cdot{\bm\nabla}\right){\bm u} &= -{\bm\nabla} p + \nu \nabla^2{\bm u}\quad\mathrm{and}\\
{\bm\nabla}\cdot{\bm u}&=0,
\end{align}
where $\uu$ is the flow velocity, $p$ is the fluid pressure divided by the fluid density $\rho_\mathrm{f}$, and $\nu$ is the kinematic viscosity of the fluid. The LBM is a solver for the Boltzmann equation which, for the present problem, runs as a quasi-compressible solver to find solutions to the above equations.

The primitive variables in the LBM are the so-called particle populations $f_i(\ux,t)$, which represent a discretized version of the  distribution function $f(\ux,\uxi,t)$:
\begin{equation}
f_i(\ux,t)\equiv f(\ux,\uxi_i,t),
\end{equation}
where the index $i=0,\cdots,q-1$ identifies the elements of a discretized velocity space $\{\uxi_i\}_{i=0}^{q-1}$. The properties of the numerical model depend among others on the choice of discrete velocities. In the present work, we choose the 27-velocity model, D3Q27 lattice \cite{shan:06}, which appears to show better numerical stability for the present problem than the commonly used D3Q19 lattice. The details of the 27 discrete velocities are stated as
\begin{equation}
\{\uxi_i\} = \left\{\frac{\delta x}{\delta t} (k{\bm e}_0+l{\bm e}_1+m{\bm e}_2),\quad k=0\cdots 2,\quad l=0\cdots 2,\quad m=0\cdots 2\right\},
\end{equation}
where $\delta x$ and $\delta t$ are the spatial and temporal resolutions respectively, and ${\bm e}_0$, ${\bm e}_1$, ${\bm e}_2$ are the Cartesian basis vectors.

The macroscopic variables, pressure $p$ and velocity $\uu$, are moments of the distribution function, and in the discrete case can be expressed through weighted sums of the particle populations:

\vspace{-5mm}
\begin{align}
p&=c_s^2\left(\sum_i f_i - \rho_0\right),\\\label{eq:macro-p}
\uu&=\frac{\sum_i \uxi_i f_i}{\sum_i f_i}.
\end{align}
In the present quasi-compressible approach, the constants $c_s$ and $\rho_0$ have a purely numerical meaning, and take the value $c_s^2=1/3\,\delta x^2/\delta t^2$ and $\rho_0=1$.

Lattice Boltzmann models usually approximate the collision term by a perturbation around an equilibrium value, described by the Maxwell-Boltzmann distribution. This function depends on the macroscopic variables, such as, velocity and pressure, and is most often approximated by a polynomial of a given degree, which is constructed in a way to reproduce the selected velocity moments exactly. In this work, we select a fourth-order equilibrium term and model the collision with the recursive-regularized approach presented in~\cite{malaspinas:15}, which is a more robust variation of the classical BGK algorithm (Bhatnagar, Gross \& Krook (1954) \cite{bhatnagar:54}). The evolution rule reads
\begin{equation}\label{eq:orestis-model}
f_i(\ux+\uxi_i\delta t, t + \delta t) = \fzero_i(p,\uu)+ \left(1-\frac{1}{\tau}\right)\fone_i,
\end{equation}
where $\fzero_i$ is the equilibrium population, expanded to fourth order with respect to the Mach number, as detailed in~\ref{sec:regularized-details}. The off-equilibrium population $\fone_i$, also detailed in~\ref{sec:regularized-details}, is projected onto a sub-space spanned by Hermite tensor polynomials, to filter out high-frequency fluctuations (see~\cite{malaspinas:15} for further details). Consequently, the under-resolved, turbulent simulations presented in this article are run without any LES-type turbulence model, as subgrid scales are filtered by the collision model. The variable $\tau$ represents the relaxation time. The Chapman--Enskog expansion (see \cite{chapman:60}) relates the relaxation time to the kinematic viscosity $\nu$ through
\begin{equation}
 \nu=c_s^2\delta t(\tau-\frac{1}{2}).
\end{equation}

One of the compelling advantages of the LBM is that the hydrodynamic stress tensor
\begin{equation}\label{eq:stress-tensor-definition}
{\bm\sigma}\equiv -p{\bm I} + \nu\left({\bm\nabla}{\bm u} + ({\bm\nabla}{\bm u})^T\right)
\end{equation}
can be evaluated directly from the primitive variables of the model $f_i$, without the need to compute any gradient:
\begin{equation}\label{eq:stress-tensor}
{\bm\sigma} = -p{\bm I} + \left(\frac{1}{2\tau}-1\right)\sum_i \uxi_i \uxi_i(f_i-\fzero_i),
\end{equation}
a property which is exploited in this work for the evaluation of the fluid forces on the solid particle surface. Here $\bm I$ is the identity matrix.

\subsection{Immersed boundaries: direct forcing approach}
An immersed boundary method (IBM) can represent the interaction of a fluid with an immersed solid, whose surface is not aligned with the Eulerian grid of the fluid. In particular, it can be used to simulate immersed moving solids, like the falling solid particles of this article, while the fluid mesh remains unchanged. The IBM gained some popularity through the approach developed by Peskin (1972) \cite{peskin:72} to study the heart valves. Since then, it has been developed further and applied to various areas. The basic idea of the IBM is that the surface of the solid object is approximated by a set of Lagrangian points, which are not required to coincide with the underlying Eulerian grid of the fluid. The fluid feels the presence of the solid by a singular force which is added to the momentum conservation equations. Regularized Dirac delta functions are used to transfer information from the Eulerian grid to the Lagrangian points, and vice versa. A very important variant of the classical IBM was proposed in \cite{fadlun:00}. This method, which is called the `direct forcing' approach, imposes a user-defined velocity at all points close to the immersed boundary. However, the no-slip boundary condition may not be effectively satisfied on each Lagrangian point due to conflicts with neighboring direct forcing terms. To solve this problem, Wang et al. (2008) \cite{wang:08} proposed the iterative `multi-direct forcing' approach, which can in principle be applied to any numerical method in CFD. The present work is based on the implementation of this framework in the LBM described in~\cite{ota:12}.

The goal of the presently used IBM is to compute a singular forcing term ${\bm h}_\mathrm{imm}$ along the fluid-solid interface and apply it as a body-force to the fluid. Let ${\bm x}$ and $\mathbf{X}_k$ denote the Eulerian grid points of the fluid and the Lagrangian points on the rigid solid surface respectively, and ${\bm U}_k$ the prescribed velocity on the Lagrangian points, and ${\bm u}^*\left({\bm x}, t + \delta t \right)$ is the fluid velocity computed by the LBM algorithm after the collision and streaming cycle. By using interpolation, the fluid velocity at the Lagrangian points is given by:

\begin{equation*}
{\bm u}^*\left(\mathbf{X}_k, t+\delta t\right) = \sum_{{\bm x}} {\bm u}^*\left({\bm x}, t + \delta t \right) W\left({\bm x}-\mathbf{X}_k\right)\delta x^3,
\end{equation*}
where the weight function

\vspace{-3mm}
\begin{equation*}
W(x, y, z) = \frac{1}{\delta x^3} w\left(\frac{x}{\delta x}\right) w\left(\frac{y}{\delta x}\right)  w\left(\frac{z}{\delta x}\right),
\end{equation*}
\vspace{3mm}
is a tensor product of the one-dimensional weight function

\begin{equation*}
    w(r) = \left\{ 
    \begin{array}{ll}
        \frac{1}{8}\left(3-2|r|+\sqrt{1+4|r|-4r^2}\right) & \text{if } |r| \leq 1, \\
        \frac{1}{8}\left(5-2|r|-\sqrt{-7+12|r|-4r^2}\right) & \text{if } 1 \leq |r| \leq 2, \\
        0 & \text{otherwise} .
    \end{array} \right.
\end{equation*}

In the following, we outline the iterative procedure of~\cite{ota:12} to compute the force term ${\bm h}_\mathrm{imm}\left({\bm x},t+\delta t\right)$, which is applied as a body-force on the Eulerian fluid grid to implement the IBM solid-fluid coupling. The subscript $l$ is used to denote the IBM iteration steps inside each IBM cycle. At step $l=0$, the body-force on the Lagrangian points is calculated as follows:
\begin{equation}\label{eq:ibm_step0}
{\bm h}_{l=0}\left({\bm X}_k,t+\delta t\right) = \frac{{\bm U}_k\left(t+\delta t\right)-{\bm u}^*\left({\bm X}_k,t+\delta t\right)}{\delta t}.
\end{equation}
Then, at step $l=1$, the body-force on the Eulerian fluid grid points is interpolated from the values of the body-force on the nearby Lagrangian surface points:
\begin{equation}\label{eq:ibm_step1}
{\bm h}_{l=1}\left({\bm x}, t+\delta t\right) = {\bm h}_\mathrm{imm}\left({\bm x},t+\delta t\right) = \sum_{{\bm X}_k} {\bm h}_{l=0}\left({\bm X}_k, t + \delta t \right) W\left({\bm x}-{\bm X}_k\right)\delta {\bm S}_k,
\end{equation}
where $\delta {\bm S}_k$ is the surface area which corresponds to the surface mesh vertex ${\bm X}_k$. At step $l=2$, the corrected velocity at the Eulerian grid points is computed as:
\begin{equation}\label{eq:ibm_step2}
{\bm u}_{l=2}\left({\bm x}, t+\delta t\right) = {\bm u}^*\left({\bm x}, t+\delta t\right) + {\bm h}_{l=1}\left({\bm x}, t+\delta t\right) \delta t.
\end{equation}
At step $l=3$, the current velocity at the Lagrangian points is interpolated:
\begin{equation}\label{eq:ibm_step3}
{\bm u}_{l=3}\left({\bm X}_k, t+\delta t\right) = \sum_{{\bm x}} {\bm u}_{l=2}\left({\bm x}, t + \delta t \right) W\left({\bm x}-{\bm X}_k\right)\delta x^3.
\end{equation}
Finally, at step $l=4$, the body-force on the solid Lagrangian points from step $l=0$ is updated:

\begin{equation*}
{\bm h}_{l=4}\left({\bm X}_k,t+\delta t\right) = {\bm h}_{l=0}\left({\bm X}_k,t+\delta t\right) + \frac{{\bm U}_k\left(t+\delta t\right)-{\bm u}_{l=3}\left({\bm X}_k,t+\delta t\right)}{\delta t}.
\end{equation*}
For the next IBM cycle, we repeat from step $l=1$. In the present work, all solid surfaces are represented by triangulated meshes. The Lagrangian points ${\bm X}_k$ correspond to the vertices of this mesh, and the surface area  $\delta {\bm S}_k$ corresponding to a vertex, is computed as one third of the sum of the areas of all triangles sharing this vertex.

Following the guidelines provided in \cite{ota:12}, we apply a total of $n=4$ IBM cycles after each iteration of the fluid solver. The immersed force is determined from the result of the last IBM cycle: ${\bm h}_\mathrm{imm}({\bm x},t+1) = {\bm h}_{\mathrm{imm},n=4}({\bm x},t+\delta t) = {\bm h}_{l=1,n=4}\left({\bm x}, t+\delta t\right)$. To apply the body-force term to the LBM algorithm, we follow the procedure proposed in~\cite{shan:06} and apply a correction to the momentum term of the equilibrium populations $\fzero$. The evolution rule of Eq.~\ref{eq:orestis-model} then updated as:
\begin{equation}\label{eq:orestis-model-with-force}
 f_i(\ux+\uxi_i\delta t, t + \delta t) = \fzero_i(p,\uu+\tau\delta t\,{\bm h}_\mathrm{imm})+ \left(1-\frac{1}{\tau}\right)\fone_i.
\end{equation}

The immersed force term ${\bm h}_\mathrm{imm}$ could be used to compute the total force and torque of the fluid, which are acting on the solid particle.
However, these forces when computed from the hydrodynamic stress tensor (defined in Eq. \ref{eq:stress-tensor-definition}), show better accuracy during numerical validation.
To compute the force $\bm F$, the stress tensor ${\bm\sigma}$ is projected onto the surface normal $\bm n$ pointing from the surface into the fluid, and the result is integrated along the surface:
\begin{equation}
\frac{{\bm F}}{\rho_\mathrm{f}} = \int_S d{\bm S} {\bm\sigma}\cdot{\bm n} \approx \sum_{{\bm X}_k} \delta{\bm S}({\bm X}_k){\bm\tau}({\bm X}_k){\bm n}({\bm X}_k).
\end{equation}
The stress tensor ${\bm\tau}$ is computed on Eulerian fluid grid according to Eq. \ref{eq:stress-tensor} and interpolated onto Lagrangian solid surface points using the weight function $W$:
\begin{equation}
\frac{{\bm F}}{\rho_\mathrm{f}} = \sum_{{\bm X}_k} \delta{\bm S}({\bm X}_k)\left(\sum_{\bm x}{\bm\tau}({\bm x})W({\bm x}-{\bm X}_k)\right) {\bm n}({\bm X}_k).
\end{equation}
The torque ${\bm T}$, relative to the center of mass ${\bm C}_m$ of the solid particle, is approximated as follows:
\begin{equation}
\frac{{\bm T}}{\rho_\mathrm{f}} = \sum_{{\bm X}_k} \delta{\bm S}({\bm X}_k)\left[\left(\sum_{\bm x}{\bm\tau}({\bm x})W({\bm x}-{\bm X}_k)\right) {\bm n}({\bm X}_k)\right]\wedge({\bm X}_k-{\bm C}_m).
\end{equation}

\subsection{Rigid particle motion}
The particles in this article are modeled as rigid bodies whose location and orientation evolve according to the equations of motion of classical mechanics. Given the total force ${\bm F}$ acting on a particle of mass $m$, the particle's center of mass ${\bm x}$ and velocity ${\bm v}$ evolve according to

\vspace{-5mm}
\begin{align}
   \frac{d{\bm v}}{dt}&= \frac{{\bm F}}{m}\quad\mathrm{and}\\
   \frac{d{\bm x}}{dt}&= {\bm v}.
\end{align}
To specify the rotational motion of the particle, we introduce an orthonormal system of vectors [${\bm d}_0, {\bm d}_1, {\bm d}_2$],
which are tied to the three Cartesian axes of the rigid particle and follow its motion. These vectors describe a local system of coordinates with origin at the particle's center of mass. The equations of motion of these unitary vectors, or of any other vector ${\bm d}$ tied to the rigid particle motion, is given by
\begin{equation}\label{eq:d-equation}
\frac{d}{dt}{\bm d} = {\bm \omega},
\end{equation}
where $\bm\omega$ is the angular velocity. The particle's moments of inertia are represented by the inertia tensor $\bm J$, which, if the mass distribution in the particle is homogeneous, is a purely geometric property:
\begin{equation}
    \frac{\bm J}{m} = \int_V d^3x\left|{\bm x}\right|^2.
\end{equation}
The time evolution of the particle's rotational degrees of freedom reads
\begin{equation}
    \frac{d}{dt}{\bm \omega} = {\bm J}^{-1}{\bm T}.
\end{equation}
To end this summary of rigid-body mechanics, we point out that the instantaneous velocity of each point $\bm x$ of the particle is expressed as
\begin{equation}
{\bm v}_{\bm x} = {\bm v} + {\bm\omega}\wedge({\bm x}-{\bm r}).
\end{equation}
This relation is used to compute the prescribed velocities ${\bm U}_k$ on the Lagrangian points ${\bm X}_k$ along the surface of the rigid body, in Steps $l=0$ (Eq.~\ref{eq:ibm_step0}) and $l=3$ (Eq.~\ref{eq:ibm_step3}) of the immersed boundary algorithm.

The particle state is updated in time steps $\delta t$ equal to those of the fluid, using an explicit time stepping scheme. To achieve second-order accuracy, a Verlet algorithm is applied, in which the position $\bm r$ and orientation ${\bm d}_i$ are defined at integral time steps $t$, while the velocity $\bm v$ and rate of rotation $\bm\omega$ are defined at half-interval steps $t-\delta t/2$. To circumvent the issue of the left-hand-side and right-hand-side terms of Eq.~\ref{eq:d-equation} being defined at different times, the updated value of the vectors ${\bm d}_i$ is first predicted as ${\bm d}_i^{\mathrm{pred}}(t+\delta t)$, and then adjusted according to a predictor-corrector scheme. Finally, to reflect the ratio of inertial effects between fluid and solid, we introduce the particle volume $V$, and the density ratio $\kappa = \rho_\mathrm{p}/\rho_\mathrm{f}$
between the mass-densities of the particle $\rho_\mathrm{p}$ and the fluid $\rho_\mathrm{f}$. The full algorithm then takes the following shape:
\begin{eqnarray}
    {\bm a(t)} &=& {\bm g} + \frac{\bm{F}(t)}{m} = {\bm g} + \frac{\bm{F}(t)}{\rho_\mathrm{f}}\frac{1}{\kappa V},\\
    {\bm v}(t+\delta t/2) &=& {\bm v}(t-\delta t/2) + \delta t*{\bm a}(t),\\
    {\bm r}(t+\delta t) &=& {\bm r}(t) + \delta t*{\bm v(t+\delta t/2)},\\
    {\bm \omega}(t+\delta t/2) &=& {\bm \omega}(t-\delta t/2) +  \delta t*\left({\bm{\mathcal{M}}}^{-1} {\bm J}^{-1}{\bm{\mathcal{M}}}\right)\cdot{\bm T}(t),\nonumber \\
                               &=& {\bm \omega}(t-\delta t/2) +  \delta t*\left({\bm{\mathcal{M}}}^{-1} \frac{1}{\kappa V} {\left(\frac{\bm{J}}{m}\right)}^{-1}{\bm{\mathcal{M}}}\right)\cdot\frac{\bm{T}(t)}{\rho_\mathrm{f}},\\
    {\bm d}_i^{\mathrm{pred}}(t+\delta t) &=& {\bm d}_i(t) + \delta t*\bm{\omega}(t+\delta t/2)\wedge\bm{d}_i(t),\\
    {\bm d}_i(t+\delta t) &=& {\bm d}_i(t) + \delta t*\bm{\omega}(t+\delta t/2)\wedge(0.5*(\bm{d}_i(t)+\bm{d}_i^{\mathrm{pred}}(t+\delta t))).
\end{eqnarray}
Here, $\bm{\mathcal{M}}$ is the transformation matrix which rotates the rigid body from its state at time $t$ to its original orientation at time $0$. If we take the orthonormal vectors ${\bm d}_0$, ${\bm d}_1$, ${\bm d}_2$ to be equal to the Cartesian basis vectors at time $0$, $\bm{\mathcal{M}}$ takes the shape of a matrix the lines of which corresponds to ${\bm d}_i(t)$.

\subsection{Coupling and time iteration}

For the time-integration of the particle motion, a leap-frog algorithm was applied in which position and velocity are represented at points on the time axis with relative shifts of half-integer time steps. The same principle is also applied to the orientation and the rotation vector. For this scheme to work properly, the successive iterations of the flow solver and the particle algorithm must be arranged appropriately, as summarised in the Table~\ref{tab:time_iteration}.

\begin{table}[!htb]
\begin{center}
\begin{tabular}{l|l|lllll}
\hline \hline
    \textbf{Event}      & \textbf{Requires}                  & $f_i$ & ${\bm\sigma}$ & ${\bm h}_\mathrm{imm}$ & ${\bm r}$, ${\bm d}_i$ & ${\bm v}$, ${\bm \omega}$ \\\hline
    (0) Initial State   &                                    & $t$   & $t$         & $t$                    & $t$   & $t-dt/2$  \\
    (1) Collide/stream  & $f_i(t),  {\bm h}_\mathrm{imm}(t)$ & $t+1$ & $t$         & $t$                    & $t$   & $t-dt/2$  \\
    (2) Advance particle    & ${\bm\sigma}(t)$, $\mathrm{particle}(t)$ & $t+1$ & $t$         & $t$                    & $t+1$ & $t+dt/2$  \\
    (3) Compute stress  & $f_i(t+1)$                         & $t+1$ & $t+1$       & $t$                    & $t+1$ & $t+dt/2$  \\
    (4) IBM iterations  & $f_i(t+1)$, $\mathrm{particle}(t+1)$   & $t+1$ & $t+1$       & $t+1$                  & $t+1$ & $t+dt/2$  \\\hline   \hline 
\end{tabular}
\caption{Sequence of events for a time iteration of the coupled fluid -- rigid-body algorithm. At each event, the second column specifies the time at which the input values are defined, and the subsequent columns the time to which the output values are taken. In the second column, the generic expression `$\mathrm{particle}$' is used to refer to the complete particle state, \textit{i.e.} to the set of variables ${\bm r}$, ${\bm d}_i$, ${\bm v}$, and ${\bm \omega}$.}\label{tab:time_iteration}
\end{center}
\end{table}

\begin{figure}[!htb]
\centering
\includegraphics[height=0.6\textwidth]{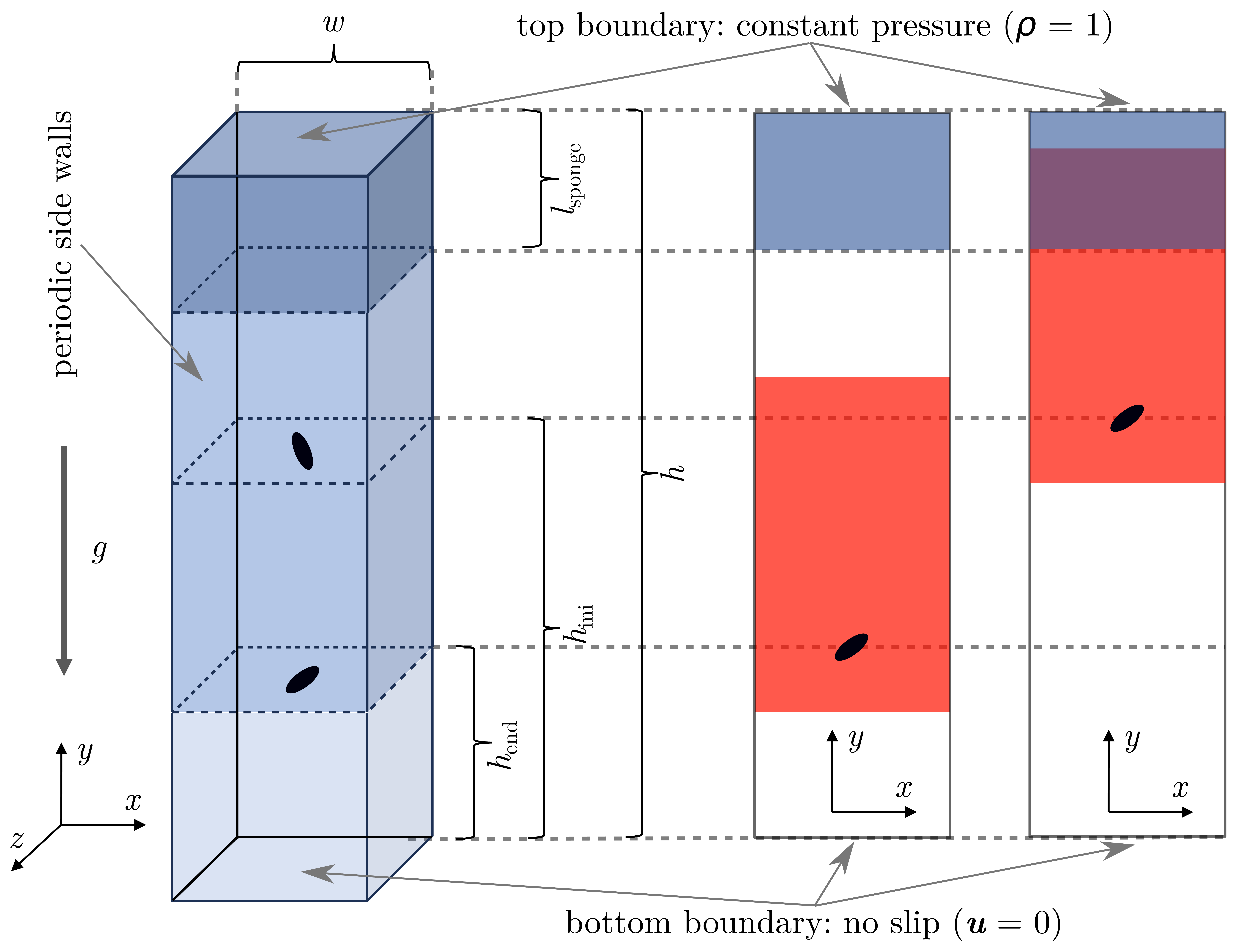}
\caption{Simulation domain: 3D computational domain for a particle (depicted as a black ellipsoid) falling in negative-$y$ direction from the release position $h_\mathrm{ini}$ to the final position $h_\mathrm{end}$ is shown in the left. The domain height is $h$ in the $y$-direction and the lateral width is $w$, and a sponge zone $l_\mathrm{sponge}$ is allocated at the top portion of the domain. Once a particle reaches its $y$-direction limit $h_\mathrm{end}$, the red portion of the domain is copied together with the particle back to $h_\mathrm{ini}$, as shown in the right. These values are specified in Table~\ref{tab:numerical_parameters} or in corresponding results subsections for different cases considered in this study. See text for further explanations. No-slip boundary condition $\bm u = 0$ and constant pressure are specified at the bottom and top boundaries, respectively. The lateral boundaries (i.e. all side walls parallel to $g$) have periodic boundary conditions. }\label{fig:freefall-setup}
\end{figure}

\subsection{Simulation setup}
The setup and simulation principles of a numerical free-fall experiment is illustrated in Fig.~\ref{fig:freefall-setup}. 
A particle of arbitrary shape is dropped either from rest or with an initial translational and / or angular velocity, and falls due to gravity in negative $y$-direction. The particle's translational and rotational vectors are free and are impacted only by gravity $g$ and by air pressure and shearing forces. The top boundary, at the maximal $y$-coordinate, implements a constant fluid density, i.e. constant-pressure boundary condition. A sponge zone is additionally defined in the vicinity of the top boundary, to eliminate spurious pressure waves created in the domain by numerical artifacts. In this zone, the fluid viscosity is increased linearly to reach a relaxation time $\tau=1$ on the top boundary. The bottom boundary implements a no-slip condition.
The lateral boundaries are periodic.

It should be noted that the distance traversed by a particle until it reaches its terminal velocity is typically several orders of magnitude larger than the particle diameter. Therefore, to avoid the allocation of an impractically large domain, the particle is periodically re-injected at its initial position together with a copy of a portion of the fluid domain around it, as illustrated in red rectangle in Fig.~\ref{fig:freefall-setup}. This copied domain contains the full fluid volume directly impacted by the particle motion, with the full width of $w$ and $\sim 60\%$ of the $h$. Also, when the particle drifts close to the lateral boundaries of the domain, a similar fluid volume around the particle is copied and re-positioned in the center of the domain. This area is typically elongated above the particle, especially in high-Reynolds flows with a fully developed wake. 

It should be pointed out that this iterative re-injection scheme cannot be replaced by a simple periodicity condition between the top and the bottom wall, as such a setup would represent an open system with ever-increasing kinetic energy, under the effect of gravity. Another setup with lateral no-slip wall (instead of a no-slip bottom) would not be acceptable either, because it would represent a motion in a channel with a cross-channel velocity profile that is not representative of particle free-fall in an unbounded fluid.

The simulation setup is characterized by the domain height $h$ and width $w$, the positions $h_\mathrm{ini}$ and $h_\mathrm{end}$ for the initial and final position of a particle (before re-injection), the length of the sponge zone $l_\mathrm{sponge}$ and the number of lattices or the resolution $N$ spatially discretizes domain width $w$. To define the particle properties, the geometric parameters of the particle, diameter $d$ of a volume-equivalent sphere, particle density $\rho_\mathrm{p}$ and the initial conditions, such as initial orientation and velocity need to be provided. The fluid properties in the simulation are defined with the fluid density $\rho_\mathrm{f}$ and its kinematic viscosity $\nu$.


\section{Results}\label{sec:results}

The numerical free-fall experiment is validated in the following subsections for spherical, spheroidal, and an irregular volcanic particles at different Reynolds numbers. The numerical parameters of the Palabos Turret simulations are summarized in Table~\ref{tab:numerical_parameters} or in corresponding results subsections for different cases considered in this study, where the parameters with * are dimensionless.
The discrete cell spacing is defined as $\delta x = w/N$. The time step $\delta t$ is defined by normalizing the terminal velocity (analytical, empirical, or experimental prediction) to a value of 0.02 in lattice units. In all simulations, gravity $g$ has a value of 9.81 m/s$^2$. The Palabos Turret simulations are performed in the parallelized super-computing facilities of the Max Planck Society. The results of the Palabos Turret simulations are compared with analytical models and experimental measurements performed with the G\"ottingen Turret \cite{bhowmick_PRL_2024} and the UNIGE Vertical Wind Tunnel \cite{Bagheri_2013, Bagheri_Bonadonna_2016}. The settling and terminal velocities presented in the results are the absolute of the particle velocity component in the $y$ direction, i.e. $|v_y|$. The terminal velocity $v_\mathrm{T}$ is defined as $|v_y|$ at the point in time at which $|a_y|$ falls below 0.01 m/s$^2$.

\subsection{Domain and mesh independence tests}

The optimization of two numerical parameters, the size of the simulation domain and  the resolution, is essential for the accurate simulation of a physical problem, e.g. to minimize the wall-effects and to reduce blockage contributions, while minimizing computational costs.
To determine the optimal lateral dimension of the simulation domain $w$, a series of domain independence tests is performed both for spherical and non-spherical particles.
$w$ is varied between 5 and 20 times the largest dimension of the simulated particle.
The vertical dimension of the domain $h$ is varied between 2 and 5 times $w$.
The discrete cell spacing $\delta x$ is varied between 1/40 to 1/12 times the smallest dimension of the simulated particle.
It is found that the wall effect of the domain boundary is evident when $w$ is less than 8 times the largest dimension of the particle.
When $h$ is less than 3$w$, the vertical dimension of the domain is too short to accommodate the long wake behind high $\mathrm{Re}$ particle.
$\delta x$ of 1/10 times the smallest particle dimension is necessary for solving the boundary layer around the particle and for convergence of terminal velocity.
Excellent agreement of the Palabos Turret simulation with analytical and experimental data confirms the choice of numerical parameters.
As a result, for all Palabos Turret simulated cases, we have ensured that $w$ is $\ge$ 10 times the largest dimension of the simulated particle, $h$ is $\ge 3w$, and $\delta x$ is $\le$ 1/10 times the smallest dimension of the simulated particle.
Obviously, for particle shapes and Reynolds numbers other than those simulated in this study, the values for setting up the computational domain may need to be optimised in order to obtain reliable results. 

\begin{figure}[!htb]
\centering
\includegraphics[width=0.9\textwidth]{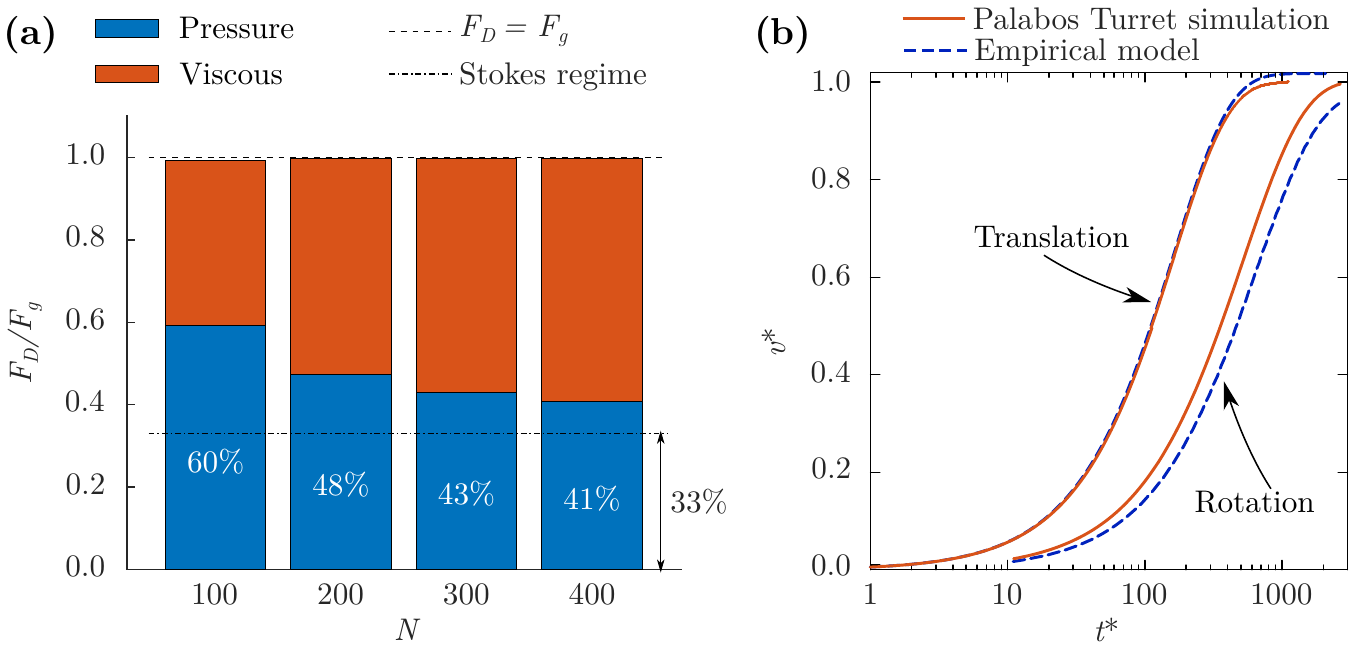}
\caption{\textbf{(a)} Impact of grid resolution $N$ on the convergence of drag force $F_D$ to the gravitational force $F_g=mg$, and its pressure and shear / viscous force contributions. Here $m$ is the mass of the particle. The values are the percentage of the pressure force to the total drag force. At the Stokes regime $\mathrm{Re}<<1$, the hydrodynamic pressure forces on the particle surface contributes 33\% to the terminal drag force \cite{Clift2005}.
\textbf{(b)} Validation of the time series of settling velocity of a sphere during free-fall (left) and free-rotation (right) with the empirical Eqs. \ref{eq:BBO} and \ref{eq:rotational}.
Time $t$ and settling velocity $v$ is non-dimensionalized as $t^*=t*v_\mathrm{T}/d$ and $v^*=v/v_\mathrm{T}$. $v_\mathrm{T}$ is the terminal velocity from the Palabos Turret simulation. The curves from the free-rotation Palabos Turret simulation is shifted in time for better visibility.
}\label{fig:convergence}
\end{figure}

A subsequent grid convergence study was conducted at a non-vanishing Reynolds number $\mathrm{Re}\sim5$, using the numerical parameters of `Sphere-1' in Table~\ref{tab:numerical_parameters}. The study demonstrates a convergence of the drag force $F_D$ on a sphere to its gravitational force $F_g$ as a function of the grid resolution (Figure~\ref{fig:convergence}(a)), and the relative difference between $F_g$ and the terminal state $F_D$ is already 0.7\% even at a low resolution of $N=100$ or $\delta x = d / 10$. A decomposition into force contributions, however, shows that the pressure force and the shear / viscous force, when considered individually, are inaccurate at low resolution and start showing signs of grid convergence at a higher resolution of $N=300$ only. 
It must be emphasized that proper numerical values of the the pressure and shear components of the force are critically important to model the free-fall behavior, and especially for the onset of rotational motion of the particles. While it can be tempting to choose a grid resolution on ground of a convergence study of the drag force alone, it is clear that this would lead to under-resolved simulations.

\begin{table}[!htb]
\begin{center}
\begin{adjustbox}{width=\textwidth}
\begin{tabular}{l|ccccc|cccccc|cc}
\hline \hline
~ &\multicolumn{5}{c}{\underline{Physical parameters}} &\multicolumn{6}{c}{\underline{Numerical parameters}} &\multicolumn{2}{c}{\underline{Results}} \\ [1ex] 
~ & $d$ [mm] &  $\kappa$ & $\nu$ [$\mathrm{m^2/s}$] &  $el$ & $fl$ & $w^*$&  $h^*$ & $h_\mathrm{ini}^*$ &  $h_\mathrm{end}^*$ & $l_\mathrm{sponge}^*$ &  $N$& $v_\mathrm{T}$ [$\mathrm{m/s}$] & $\mathrm{Re}$\\
\hline
Sphere-1 & 0.148 & 996.26 & 1.51e-5 & 1.0 & 1.0 & 10.0 & 40.0 & 35.0 & 5.0 & 4.0 & 400 & 0.5233   & 5.12 \\
Sphere-2 & 0.140 & 996.26 & 1.51e-5 & 1.0 & 1.0 & 10.0 & 40.0 & 35.0 & 5.0 & 4.0 & 300 & 0.4848 & 4.49 \\
Spheroid-1 & 0.140 & 996.26 & 1.51e-5 & 1.0 & 0.2 & 17.1 & 51.3 & 46.2 & 5.1 & 4.1 & 550 & 0.3405 & 3.15\\
Spheroid-2 & 0.140 & 996.26 & 1.51e-5 & 0.2 & 1.0 & 29.3 & 87.9 & 79.1 & 8.8 & 7.0 & 550 & 0.3630 & 3.36 \\
\hline \hline
\end{tabular}
\end{adjustbox}
\end{center}
\caption{Simulation parameters and a summary of results of the virtual free-fall Palabos Turret simulations. $d$ is the diameter of a volume-equivalent sphere, $\kappa=\rho_\mathrm{p}/\rho_\mathrm{f}$ is the particle to fluid density ratio, $\nu$ is the kinematic viscosity of fluid, and particle shape parameters, such as elongation $el = b/a$ and flatness $fl = c/b$ are defined as the ratios of the largest $a$, intermediate $b$ and smallest $c$ dimensions of the particle, orthogonal to each other. Numerical length parameters are non-dimensionalized with respect to $d$, e.g. $w^* = w / d$, $h^* = h / d$. For simulations with multiple resolution for the same particle, the value of resolution $N$, terminal velocity $v_\mathrm{T}$, and particle Reynolds number $\mathrm{Re}$ here correspond to the simulation with the highest resolution.
The last two columns show the average $v_\mathrm{T}$ in the statistically stationary state, as well as the associated $\mathrm{Re}$.}
\label{tab:numerical_parameters}
\end{table}

\subsection{Transient motion of a sphere at low Reynolds number}

For the validation of sphere motion in inertial flow at different Reynolds numbers, we compare it to the empirical relationship proposed by Clift \& Gauvin (1971) \cite{clift1971motion}.
This model presents the standard drag curve throughout the transitional and Newton regimes ($\mathrm{Re} < 10^5$):
\begin{equation}\label{eq:CG71_drag_eq}
C_D = (24/\mathrm{Re}) (1 + 0.15\,\mathrm{Re}^{0.687}) + 0.42 / (1 + 4.25 \cdot 10^4\,\mathrm{Re}^{-1.16}).
\end{equation}
Through time-integration of Newton's equations of motion, this drag curve translates into a free-fall trajectory of a spherical particle from rest to terminal velocity $v_\mathrm{T}$. The equation is solved through an implicit Euler scheme, where the time evolution of the sphere velocity is given by the Basset-Boussinesque-Oseen equation \ref{eq:BBO} \citep{basset-boussinesq-oseen}. For validating the time history of settling velocity from the free-fall Palabos Turret simulation, a Reynolds of $Re=4.49$ is chosen, with parameters summarized under the label `Sphere-2' in Table~\ref{tab:numerical_parameters}. Figure~\ref{fig:convergence}(b) compares the obtained numerical result and empirical curve from Eqs. \ref{eq:CG71_drag_eq} and \ref{eq:BBO}. The time series of the free-fall settling velocity from the Palabos Turret simulation is in excellent agreement with the empirical curve, and shows a relative difference of 1.7\% when comparing the terminal velocities.

For the validation of sphere motion in a shear flow with no gravity, we simulated a sphere that starts to rotate without external torque under the effect of a linear shear flow in the ambient. The far-field flow velocity is defined as
\begin{equation*}
    v_x = 0, \quad v_y = v_0 + Gy, \quad v_z=0,
\end{equation*}
where the coordinates $x$, $y$, and $z$ represent a fixed reference frame. Here, $G$ is the shear rate of the ambient flow, and we denote the rotation rate of the ambient flow as $\omega_f=G/2$. An analytical formulation for the rational motion of a sphere in inertial flow is detailed in \ref{sec:equationsAppendix}. As pointed out for example in~\citep{lin70}, the terminal angular velocity of a sphere is equal to the ambient flow rate in a Stokes flow, $\omega_f=\omega_\mathrm{st}$, while a slip angular velocity is observed at finite Reynolds numbers. We define the Reynolds number in a rotational flow as $\mathrm{Re}_G = 4Gr^2/\nu$,
where $r$ is the radius of the sphere. The torque term in Eq.~\ref{eq:rotational} is therefore $T = 9\mathrm{Re}_Gr\pi\nu^2\rho_\mathrm{f}/16$.
In the Palabos Turret simulation, the Reynolds number is fixed to a finite value of $\mathrm{Re}\sim 5$, with simulation parameters as $d = 0.140$ mm, $\kappa = 1000$, $\nu=$ 1.73e-5 m$^2$/s, $w^*=20$, $h^*=30$, $h_\mathrm{ini}^*=20$, $h_\mathrm{end}^*=10$, $l_\mathrm{sponge}^*=3$, and $N=300$. For the rotational motion, 
the gravity is disabled, and a no-slip boundary condition without sponge zone is applied on all walls. The numerical result of rotational motion is shown in Figure \ref{fig:convergence}(b), and compared to the empirical model Eq. \ref{eq:rotational}. The time series of the free-rotation settling velocity from the Palabos Turret simulation is in good agreement with the empirical curve despite the sphere is simulated at a Reynolds number beyond the Stokes regime.
The relative difference in the terminal velocities from the Palabos Turret simulation and the analytical model is 4\%.

\subsection{Terminal velocities of spheres at different Reynolds number}\label{sec:validation2}

For the next set of validation, we compared the terminal velocities $v_\mathrm{T}$ of freely falling spheres of different sizes, thus different Reynolds numbers.
For this case, heavy spheres of density $\rho_\mathrm{p} = 1200$ kg/m$^3$ are simulated to fall in air of density $\rho_\mathrm{f} = 1.2045$ kg/m$^3$, resulting to a high particle to fluid density ratio.
Such simulations are more challenging than a simple numerical evaluation of the sphere drag.
At such high density ratio, these spheres exhibit higher settling velocity and the algorithm needs to accommodate the rotational and translational motion of the solid body in the fluid grid.
Moreover, with an increase in sphere Reynolds number $\mathrm{Re}$, the normalized vertical free-fall distance to reach the terminal state increases exponentially.
This results in an exponential increase in the computational time to reach the terminal velocity at higher $\mathrm{Re}$.

The Palabos Turret simulation domain boundaries for this validation had a fixed extent of $w^*=10$ and $h^*=30$.
However, depending on the $\mathrm{Re}$, $h_\mathrm{ini}^*$, $h_\mathrm{end}^*$ and $l_\mathrm{sponge}^*$ are varied to accommodate the gradually increasing wake region behind the sphere as the Reynolds number increases.
When $\mathrm{Re}<30$, $h_\mathrm{ini}^* = 27.0$, $h_\mathrm{end}^*=3.0$ and $l_\mathrm{sponge}^*=2.4$ are used with a grid resolution of $N=300$.
When $20\le\mathrm{Re}<1000$, $h_\mathrm{ini}^* = 23.0$, $h_\mathrm{end}^*=3.0$ and $l_\mathrm{sponge}^*=2.4$ are used with the same resolution of $N=300$.
And when $\mathrm{Re}\ge1000$, $h_\mathrm{ini}^* = 20.0$, $h_\mathrm{end}^*=3.0$ and $l_\mathrm{sponge}^*=2.4$ are used with an increased resolution of $N=400$ to accommodate for the thinning of the boundary layers near the sphere surface and for capturing the small scale vortical structures in the wake region.

\begin{table}[!htb]
\begin{center}
\begin{tabular}{l|l|l|l|l|l|l|l|l|l|l}
\hline \hline
$d$ [mm] & 0.062 & 0.088 & 0.140 & 0.148 & 0.160 & 0.315 & 0.659 & 1.488 & 3.770 & 10.530 \\
$v_\mathrm{T}^D$ [m/s] & 0.142 & 0.239 & 0.485 & 0.524 & 0.584 & 1.358 & 2.933 & 6.035 & 10.29 & 17.394 \\
\hline
$\mathrm{Re}$ & 0.58 & 1.39 & 4.49 & 5.13 & 6.18 & 28.30 & 127.88 & 594.13 & 2566.6 & 12118 \\
\hline
$\Delta F$ [\%] & 0.01 & 0.24 & 0.28 & 0.19 & 0.48 & 0.72 & 0.96 & 1.04 & 1.48 & 3.12 \\
$v_\mathrm{T}^R$ [m/s] & 0.138 & 0.237 & 0.488 & 0.529 & 0.590 & 1.382 & 2.952 & 5.993 & 10.891 & 17.736 \\
$\Delta v_\mathrm{T}$ [\%] & 2.82 & 0.69 & -0.71 & -0.91 & -1.09 & -1.75 & -0.65 & 0.70 & -5.84 & -1.97 \\
$p^*$ [\%] & 35.80 & 38.76 & 42.67 & 43.10 & 43.46 & 48.37 & 57.50 & 68.90 & 88.20 & 97.09 \\
\hline \hline
\end{tabular}
\caption{Evolution of the terminal velocities of freely falling spheres for $5.8\times10^{-1}<\mathrm{Re}<1.2\times10^{4}$.
The terminal velocity $v_\mathrm{T}^D$ obtained from the Palabos Turret simulation is used to calculate particle Reynolds number as $\mathrm{Re} = dv_\mathrm{T}^D/\nu$.
The kinematic viscosity of the fluid $\nu$ is $1.51\times 10^{-5}$ m$^2$/s.
The relative difference of the gravitational force $F_g=mg$ and terminal state drag force $F_D$ from the Palabos Turret simulation is shown as $\Delta F = (F_g-F_D)/F_g$.
Based on $\mathrm{Re}$, a empirical model \cite{clift1971motion} is used to approximate the drag coefficient $C_D$ (Eq. \ref{eq:CG71_drag_eq}), and a reference terminal velocity is obtained as $v_\mathrm{T}^R = (8F_g/(\pi d^2 \rho_\mathrm{f} C_D))^{0.5}$.
The density of the air $\rho_\mathrm{f}$ is $1.2045$ kg/m$^3$, where $\nu$ and $\rho_\mathrm{f}$ are representative of the air at 20$^o$C at the mean sea level.
For the low $\mathrm{Re}$ case ($\mathrm{Re}=0.58$), the $v_\mathrm{T}^R$ is computed from Stokes' law as $\rho_\mathrm{p} gd^2/(18 \rho_\mathrm{f} \nu)$.
The density of the sphere $\rho_\mathrm{p}$ is $1200$ kg/m$^3$.
The relative difference of velocities is $\Delta v_\mathrm{T} = (v_\mathrm{T}^D - v_\mathrm{T}^R)/v_\mathrm{T}^D$.
$p^*$ indicate the contribution of the pressure force to the total drag force at the terminal state obtained from the stress tensor computed in the Palabos Turret simulations.
}\label{tab:terminal_velocity}
\end{center}
\end{table}

\begin{figure}[htb]
\centering
\includegraphics[width=\textwidth]{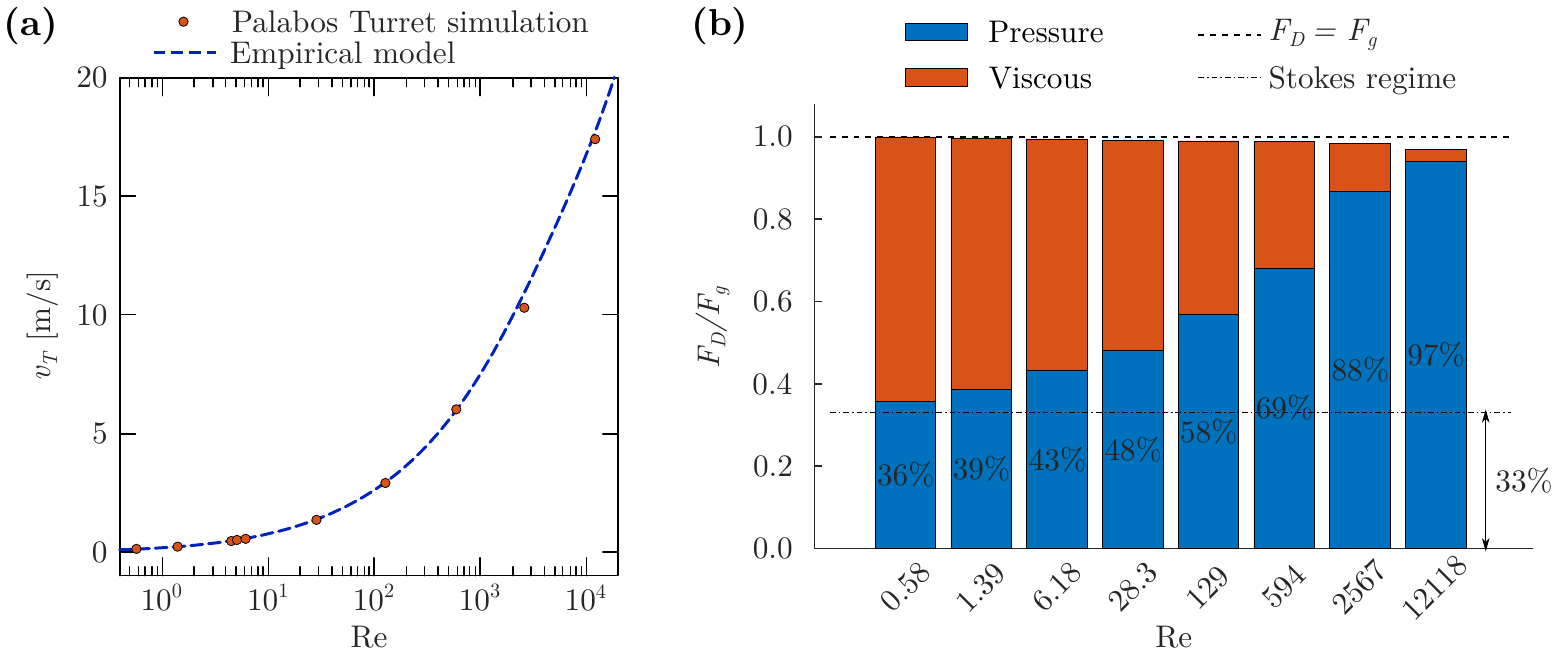}
\caption{\textbf{(a)} Comparison of terminal velocities $v_\mathrm{T}$ of falling spheres at various Reynolds number $\mathrm{Re}$ between the Palabos Turret simulation (Table~\ref{tab:terminal_velocity}) and an empirical drag model (Eq. \ref{eq:CG71_drag_eq}).
For this empirical terminal velocity calculation, the terminal state drag force is considered to be equal to the gravitational force on the sphere.
\textbf{(b)} Evolution of the terminal state pressure and shear / viscous force contribution in the total drag force for various $\mathrm{Re}$. The values are the percentage of the pressure force to the total drag force, shown as $p^*$ in Table~\ref{tab:terminal_velocity}.
}\label{fig:terminal_velocity}
\end{figure}

The results are summarized in Table~\ref{tab:terminal_velocity} and Figure~\ref{fig:terminal_velocity}.
It can be seen that in a laminar regime, the numerical terminal velocity matches the empirical value with a two to three-digit accuracy (Figure~\ref{fig:terminal_velocity}(a)).
The terminal state drag force $F_D$ obtained from the Palabos Turret simulation in this regime has less than 0.5\% relative difference from the gravitational force.
For the higher Reynolds number up to order of 10000, the Palabos Turret simulation results of the terminal state drag force show a relative difference up to 3\% from the gravitational force (Figure~\ref{fig:terminal_velocity}(b)), despite the increase in resolution $N$ from 300 to 400 for the $\mathrm{Re}\ge1000$ cases.
This results in $\sim137\%$ increase in mesh size, from 81 million to 192 million lattices.
To improve the accuracy further, a finer mesh resolution is required to resolve the thinning boundary layer well, but at a much higher computational cost.
The terminal velocity from the Palabos Turret simulation $v_\mathrm{T}^D$ remains below $6\%$ relative difference from its empirical value $v_\mathrm{T}^R$, which itself has an error up to 6\% from the experimental values \cite{Bagheri_Bonadonna_2016}.

The comparison of the $v_\mathrm{T}^D$ for the $d=0.140$ mm sphere with the terminal velocity obtained from laboratory experiments shows a relative difference lower than 0.2\% \cite{Bhowmick_2024_Arxiv}, which further validates the Palabos Turret simulation.
In such Palabos Turret simulation, the evolution of the terminal state pressure and shear / viscous force contribution in the total drag force is well resolved and is plotted as a function of the particle Reynolds number in Figure~\ref{fig:terminal_velocity}(b).
It is observed that the pressure contribution to the drag force increases from close to 1/3 in the Stokes regime \cite{Clift2005} for $\mathrm{Re}=0.6$ to 97\% for $\mathrm{Re}=12000$.

\subsection{Terminal velocities of spheroids and transient oscillations}

For the validation of angular motion of non-spherical particles at finite Reynolds number, the next set of Palabos Turret simulations investigate the freely falling settling behavior of spheroidal particles. The naming adopted is based on the three orthogonal axes $a$, $b$, $c$ of the spheroids, defining shape parameters: $el$ for the elongation $el = b/a$ and $fl$ for the flatness $fl = c/b$. The two investigated cases are of an oblate spheroid (`Spheroid-1', $el=1.00$, $fl=0.20$) and a prolate spheroid (`Spheroid-2', $el=0.20$, $fl=1.00$), with detailed parameters provided in Table~\ref{tab:numerical_parameters}. The numerical output is compared to experimental results of settling particle behavior in the G\"ottingen Turret settling chamber~\cite{Bhowmick_2024_Arxiv}.

In this settling chamber experimental setup \cite{bhowmick_PRL_2024}, particles of different shapes with at least one dimension larger than 100 $\mathrm{\mu}$m are tracked three-dimensionally using four high-speed cameras.
The vertical distance of free-fall in still air is tracked for a vertical distance of $\sim$60 mm.
In this experimental setup, it is observed that the heavy ellipsoidal particles undergo transient oscillations to reach their terminal state \cite{bhowmick_PRL_2024,Bhowmick_2024_Arxiv} and that the microplastic fibers exhibit orientation fluctuations depending on their shape and can also have a preferred terminal orientation \cite{Tatsii_2023,candelier2024}.

Experimental and numerical observations show that spheroidal particles adopt a preferred orientation on their free-fall trajectory, with the side of largest cross-section facing the direction of gravity. If a particle is not originally aligned in this preferred direction, it oscillates around and gradually approaches the preferred orientation. To validate the numerical statistics of the freely falling behavior of spheroidal particle, we compared both the terminal velocity and the oscillation frequency with the experimental observations, as shown in Table~\ref{tab:spheroids}. An excellent match is observed for both quantities, with a relative difference within 1\% for the terminal velocity and within 2\% for the oscillation frequency.

\begin{table}[htb!]
\centering
\begin{tabular}{c|cc|ccc|ccc}
\hline \hline
& $el$   & $fl$   & $v_\mathrm{T}^D$ & $\langle V_\mathrm{T}^E\rangle$ & $\sigma (v_\mathrm{T}^E)$ & $f^D$ & $\langle f^E\rangle$ & $\sigma (f^E)$ \\ \hline
Spheroid-1 (oblate) & 1.00 & 0.20 & 0.3405 & 0.3371 & 0.0018 & 40.57 & 42.02 & 0.00 \\ 
Spheroid-2 (prolate) & 0.20 & 1.00 & 0.3630 & 0.3519 & 0.0019 & 25.57 & 27.85 & 2.03 \\ \hline \hline
\end{tabular}
\caption{Comparison of the numerical and experimental observation of the settling behavior of non-spherical spheroidal particles.
The terminal velocity and the frequency of angular oscillations of the particles from the Palabos Turret simulation are shown as $v_\mathrm{T}^D$ and $f^D$ respectively.
The mean values of the terminal velocity and the frequency of angular oscillations from the experiments are shown as $\langle V_\mathrm{T}^E\rangle$ and $\langle f^E\rangle$, and their standard deviation as $\sigma (v_\mathrm{T}^E)$ and $\sigma (f^E)$ respectively.
All quantities of the velocity and frequency are in units of m/s and Hz respectively.
}\label{tab:spheroids}
\end{table}

\subsection{Free fall of a volcanic particle of non-spherical irregular shape}

\begin{figure}[htb!]
\centering
\includegraphics[width=\textwidth]{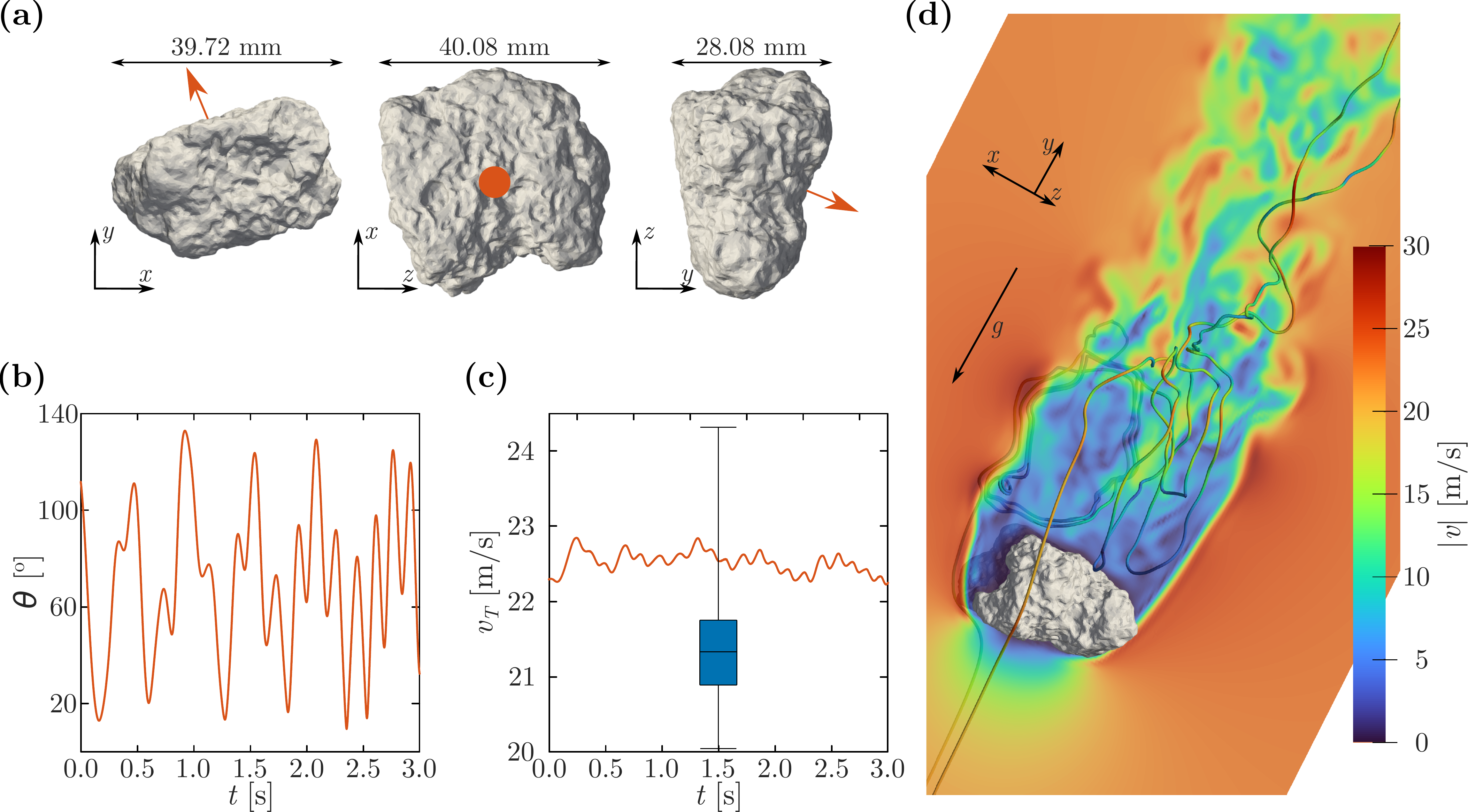}
\caption{
\textbf{The Palabos Turret simulation of an irregular particle at \bm{$Re \sim 4\times 10^4$}. (a)} 3D model of a volcanic particle captured using a 3D-scanner \cite{bagheri2015characterization} and a reference normal vector represented with an orange arrow.
\textbf{(b)} Time evolution of the angle of the reference normal vector $\theta$ - the angle between the normal vector and the perpendicular $y$ direction parallel to $g$.
\textbf{(c)} Time evolution of the settling velocity of the particle from the Palabos Turret simulation in orange solid line and a box plot of the terminal velocity of the particle $v_\mathrm{T}^E$ from experiment.
\textbf{(d)} Contours of the magnitude of fluid velocity $|v|$ along a vertical plane passing through the particle.
Two streamlines, colored also with $|v|$, captures the complex recirculating nature of the wake.
}\label{fig:particle-lapilli}
\end{figure}

The final validation case considers a specific volcanic particle of the category of lapilli, which is collected from a field campaign near a volcanic eruption.
This particle was imaged with laser line scans under a microscope, and the surface contour points were digitized and converted to a triangulated surface mesh.
This surface mesh served both as the input of particle geometry in the Palabos Turret simulation, and as a model for 3D-printing of identical particles for reproducible experimental investigations.
This case is chosen both to illustrate the capability of this presented Palabos Turret to simulate particles of arbitrary shape and to produce robust results in a turbulent regime.
The dimensions of this particle are 39.72 mm, 28.08 mm and 40.08 mm along the three orthogonal directions as shown in Figure~\ref{fig:particle-lapilli}(a), with a volume of 1.68e4 mm$^3$ which is equivalent to a sphere of diameter $d_{eq}=$ 31.8 mm.

This particle is too large to do laboratory experiment on its free-fall, because of the challenges of reproducing free-fall conditions for high Reynolds number in a laboratory setting.
A vertical wind-tunnel is not suitable, especially because of the turbulence introduced through the air inlet, which would unsettle the orientation of the particle.
Similarly to the falling behavior of non-spherical particles at lower Reynolds numbers, this particle also exhibits oscillatory and spinning motion in its falling orientation.
Such behavior dictates the free-fall statistics, including the average terminal velocity of the particle.
A realistic representation of ambient aerodynamic condition is therefore essential.

The experiment was conducted in an exterior setting involving a free-fall over a vertical distance in the order of 30 m.
Such large vertical distance allowed to measure the average terminal velocity, but without access to additional small scale statistics.
This emphasizes the need of a numerical simulation framework to complement the experimental observation and to provide insight into the details of particle motion.
The average terminal velocity obtained experimentally for this particle is $\langle v_\mathrm{T}^E\rangle=$ 21.36 m/s and it is observed that this particle undergoes rotational motion when settling under gravity.
The particle Reynolds number $\mathrm{Re} = \langle v_\mathrm{T}^E\rangle d_{eq} / \nu$ is $\sim$40000, where the kinematic viscosity of air $\nu$ is 1.73e-5 m$^2$/s.

The Palabos Turret simulation is conducted with parameters $\rho_\mathrm{f}=$ 1.096 kg/m$^3$, $\rho_\mathrm{p}=$ 1127 kg/m$^3$, $\kappa = \rho_\mathrm{p}/\rho_\mathrm{f} = 1028.28$, $w^*=12.58$, $h^*=37.74$, $h_\mathrm{ini}^*=25.16$, $h_\mathrm{end}^*=3.77$, $l_\mathrm{sponge}^*=3.02$, and $N=450$.
The numerical domain has 273.4 million lattices and was simulated with 2048 MPI (message passing interface) tasks.
The time history of the free fall of this particle consists of fluctuations in its orientation, which are shown in Figure~\ref{fig:particle-lapilli}(b) as a variation of the angle $\theta$ of the normal vector to the vertical direction.
Such angular motion changes the projection area of the particle in the falling direction, and the settling velocity of the particle therefore changes with time, as shown in Figure~\ref{fig:particle-lapilli}(c).
During the shown interval of 3 seconds, the average value of the settling velocity -- the terminal velocity $\langle v_\mathrm{T}^D\rangle$ is 22.55 m/s, which differs by 5.3\% from the experimental mean value of 21.36 m/s.
This comparison shows a good match between the Palabos Turret simulation and experiment, while the simulation reveals further insights into the angular orientations adopted by the particle, and their impact on the settling velocity.
Figure~\ref{fig:particle-lapilli}(d) shows the details of airflow around the particle, as it freely falls under the gravity.
The Palabos Turret simulation resolves the details of the boundary layer and the small scale vortices in the recirculating wake behind the particle.



\section{Conclusion} \label{sec:conclusion}

This paper presents a state-of-the-art numerical framework, the Palabos Turret, for studying the falling behaviour of particles of arbitrary shape over a wide range of particle Reynolds numbers spanning 5 orders of magnitude -- from the Stokes' to the Newton’s regime.
The recursive-regularized LBM, IBM, and a time-staggered predictor-corrector method are employed to resolve the fluid field, fluid-particle coupling, and particle motion, respectively.
A dynamic memory allocation scheme that takes into account the inherent multi-scale nature of the problem allows the simulation of any desired fall distance.
Here we have verified the excellent performance of the Palabos Turret with both analytical and experimental data sets. 

Although the translational motion of the particles can be correctly captured at fairly coarse grid resolutions, we have found that the surface stresses and rotational dynamics can only be truly captured at fine grid resolutions. For instance, integral metrics such as total drag can converge at lower resolutions, but the accurate force distribution across a particle surface (which is needed to determine the torques) requires much higher resolution. Furthermore, the relative contribution of pressure and shear terms to the force can vary with increasing resolution, even if the total force remains fairly constant. Another interesting observation is that the correct representation of the rotational dynamics of the particles is of critical importance, since the instantaneous settling velocity is a function of particle orientation. Accurate angular motion is required for the actual derivation of the statistics of the settling velocity and lateral movement. 

Now that we have verified the reliability of the Palabos Turret, it can be used as an invaluable tool to complement experimental data and to overcome the limitations of experiments in which capturing flow features, surface loads and long particle trajectories are a major challenge. 
The first application of the Palabos Turret and thorough investigations of the free fall dynamics of ellipsoids at relatively low Reynolds numbers of the order of unity are presented in~\cite{Bhowmick_2024_Arxiv}, where further excellent agreement between the Palabos Turret and experimental data is unveiled.

\section*{Acknowledgments}
T. B. was funded by the German Research Foundation (DFG) Walter Benjamin Position (Project No. 463393443). This work was supported by the Max Planck Society and University of Geneva. Scientific activities are carried out in the Max Planck Institute for Dynamics and Self-Organization (MPIDS) and in University of Geneva. Computational resources from the Max Planck Computing and Data Facility and the MPIDS are gratefully acknowledged. We thank Francesco Marson, Orestis Malaspinas and Eberhard Bodenschatz for support.


\section*{Author contributions}
Conceptualization and research design: T.B., J.L., Y.W. and G.B.; experiments: T.B.; postprocessing of the experimental data: T.B. and G.B.; simulations: T.B., J.L. and Y.W.; postprocessing of the simulation data: T.B., J.L. and Y.W.; formal analysis: T.B., J.L., Y.W. and G.B.; development of simulation software: T.B., J.L. and Y.W.; visualization: T.B., J.L. and Y.W.; validation: T.B., J.L. and Y.W.; interpretation of results: T.B., J.L., Y.W. and G.B.; writing-original draft: T.B., J.L., Y.W. and G.B; writing-review and editing: T.B., J.L., Y.W. and G.B.

\section*{Competing interests}
There are no competing interests to declare.

\section*{Data and materials availability:}
Details of the experiments are available in 
\url{https://doi.org/10.48550/arXiv.2408.11487}. Other data related to this paper may be requested from the authors.

\newpage
\appendix

\section{Regularized LBM: equilibrium and off-equilibrium}\label{sec:regularized-details}

The equilibrium and off-equilibrium populations for the regularized lattice Boltzmann method are defined as follows:
\begin{align}
  \fzero_i=&w_i(p/c_s^2+\rho_0)\left(1+\frac{\uxi_i\cdot\uu}{c_s^2}+\frac{1}{2c_s^4}\H^{(2)}_i:\uu\uu\right.\nonumber\\
	  &\quad\quad+\frac{1}{2c_s^6}\left(\H^{(3)}_{ixxy}u_x^2u_y+\H^{(3)}_{ixxz}u_x^2u_z+\H^{(3)}_{ixyy}u_xu_y^2+\H^{(3)}_{ixzz}u_xu_z^2\right.\nonumber\\
	  &\quad\quad\quad\quad\quad\quad\left.+\H^{(3)}_{iyzz}u_yu_z^2+\H^{(3)}_{iyyz}u_y^2u_z+2\H^{(3)}_{ixyz}u_xu_yu_z\right)\nonumber\\
	  &\quad\quad+\frac{1}{4c_s^8}\left(\H^{(4)}_{ixxyy}u_x^2u_y^2+\H^{(4)}_{ixxzz}u_x^2u_z^2+\H^{(4)}_{iyyzz}u_y^2u_z^2\right.\nonumber\\
	  &\quad\quad\quad\quad\quad\quad\left.+2\left(\H^{(4)}_{ixyzz}u_xu_yu_z^2+\H^{(4)}_{ixyyz}u_xu_y^2u_z+\H^{(4)}_{ixxyz}u_x^2u_yu_z\right)\right)\nonumber\\
	  &\quad\quad+\frac{1}{4c_s^{10}}\left(\H^{(5)}_{ixxyzz}u_x^2u_yu_z^2+\H^{(5)}_{ixxyyz}u_x^2u_y^2u_z+\H^{(5)}_{ixyyzz}u_xu_y^2u_z^2\right)\nonumber\\
	  &\quad\quad\left.+\frac{1}{8c_s^{12}}\H^{(6)}_{ixxyyzz}u_x^2u_y^2u_z^2\right),\label{eq_fzero_d3q27}\\
  \fone_i=&w_i\left(\frac{1}{2c_s^4}\H^{(2)}_i:\uP^{(1)}\right.\nonumber\\
	  &\quad\quad+\frac{1}{2c_s^6}\left(\H^{(3)}_{ixxy}a^{(3)}_{1,xxy}+\H^{(3)}_{ixxz}a^{(3)}_{1,xxz}+\H^{(3)}_{ixyy}a^{(3)}_{1,xyy}+\H^{(3)}_{ixzz}a^{(3)}_{1,xzz}\right.\nonumber\\
	  &\quad\quad\quad\quad\quad\quad\left.+\H^{(3)}_{iyzz}a^{(3)}_{1,yzz}+\H^{(3)}_{iyyz}a^{(3)}_{1,yyz}+2\H^{(3)}_{ixyz}a^{(3)}_{1,xyz}\right)\nonumber\\
	  &\quad\quad+\frac{1}{4c_s^8}\left(\H^{(4)}_{ixxyy}a^{(4)}_{1,xxyy}+\H^{(4)}_{ixxzz}a^{(4)}_{1,xxzz}+\H^{(4)}_{iyyzz}a^{(4)}_{1,yyzz}\right.\nonumber\\
	  &\quad\quad\quad\quad\quad\quad\left.+2\left(\H^{(4)}_{ixyzz}a^{(4)}_{1,xyzz}+\H^{(4)}_{ixyyz}a^{(4)}_{1,xyyz}+\H^{(4)}_{ixxyz}a^{(4)}_{1,xxyz}\right)\right)\nonumber\\
	  &\quad\quad+\frac{1}{4c_s^{10}}\left(\H^{(5)}_{ixxyzz}a^{(5)}_{1,xxyzz}+\H^{(5)}_{ixxyyz}a^{(5)}_{1,xxyyz}+\H^{(5)}_{ixyyzz}a^{(5)}_{1,xyyzz}\right)\nonumber\\
	  &\quad\quad\left.+\frac{1}{8c_s^{12}}\H^{(6)}_{ixxyyzz}a^{(6)}_{1,xxyyzz}\right),\label{eq_fone_d3q27}
\end{align}
$\H_i^{(n)}$ is the Hermite polynomial of order $n$. Finally $\ua_1^{(n)}$ is given by
\begin{align}
 a^{(n)}_{1,\al_1...\al_n}&=a^{(n-1)}_{1,\al_1...\al_{n-1}}u_{\al_n}+\left(u_{\al_1}...u_{\al_{n-2}}P^{(1)}_{\al_{n-1}\al_n}+\hbox{perm}(\al_n)\right)\label{eq_rel_a1}\quad\mathrm{and}\\
 \uP^{(1)}&=\sum_i \uxi_i\uxi_i (f_i-\fzero_i),
\end{align}
where $\hbox{perm}(\al_n)$ are the cyclic index permutations of indexes from $\al_1$ to $\al_{n-1}$ ($\al_n$ is never permuted). 

\section{Equations for sphere motion}\label{sec:equationsAppendix}

The Basset-Boussinesque-Oseen equation~\citep{basset-boussinesq-oseen}:
\begin{equation}
    \frac{du}{dt} = \frac{1}{\rho+\frac{1}{2}}\left(\kappa g -\frac{9}{2}\frac{v\nu}{r^2}-\frac{9}{2}\frac{\nu}{r\pi}\sqrt{\frac{\pi}{\nu}}\int_0^t\frac{du}{dt'}\frac{dt'}{\sqrt{t-t'}}\right).\label{eq:BBO}
\end{equation}
Here $r$ is the radius of the sphere.

A solution for rotational motion was first considered by Basset (1888) \citep{basset1888} and later improved by Feuillebois \& Lasek (1978) \citep{feuillebois78}. Here, a constant torque $T$ is applied to a sphere initially at rest. The sphere begins to accelerate angularly and ultimately reaches the terminal angular velocity $\omega$, given by $\omega = T\kappa/(6 m \nu)$, where $m$ is the mass of the sphere and $\kappa$ is the density ratio between the sphere and the fluid. The time evolution of the angular velocity $\omega$ is described by the relation:
\begin{align}
    \frac{d\omega}{dt} &= \frac{\rho_\mathrm{f}}{I}\left[\frac{T}{\rho_\mathrm{f}}-8\pi\nu r^3\omega-8\pi\nu r^3\left(\frac{r}{3\sqrt{\pi\nu}}\int_0^t\frac{d\omega}{dt'}\frac{dt'}{\sqrt{t-t'}}\right)\right.\label{eq:rotational}\\
                         &\quad\left.+\frac{8}{3}\pi\nu r^3\left(\int_0^t\frac{d\omega}{dt'}\exp(\nu(t-t')/r^2)\mathrm{erfc}\sqrt{\nu(t-t')/r^2}dt'\right)\right].\nonumber
\end{align}
For a sphere with homogeneous mass distribution, the moment of inertia $I$ takes the value $I=2mr^2/5$, and the term $\rho_\mathrm{f}/I$ in the equation above is given by $\rho_\mathrm{f}/I = 15/(8\kappa r^5\pi)$.

\newpage
\bibliographystyle{elsarticle-num-names}
\bibliography{JCP_biblio}

\end{document}